\documentclass[prl,aps,twocolumn]{revtex4}%

\usepackage{amssymb}
\usepackage{graphicx}

\usepackage{subfigure}

\usepackage{amsmath}

\def\eqref#1{\mbox{(\ref{#1})}}

\begin{document}
\title{Security of Power Packet Dispatching Using Differential Chaos Shift Keying}

\author{Yanzi Zhou} \email{y-zhou@dove.kuee.kyoto-u.ac.jp} 
\affiliation{Department of Electrical Engineering, Kyoto University, Katsura, Nishikyo, Kyoto, 615-8510 Japan}
\author{Ryo Takahashi} \email{takahashi.ryo.2n@kyoto-u.ac.jp}
\affiliation{Department of Electrical Engineering, Kyoto University, Katsura, Nishikyo, Kyoto, 615-8510 Japan}
\author{Takashi Hikihara} \email{hikihara.takashi.2n@kyoto-u.ac.jp}
\affiliation{Department of Electrical Engineering, Kyoto University, Katsura, Nishikyo, Kyoto, 615-8510 Japan}

\date{\today}

\begin{abstract}
This paper investigates and confirms one advantageous function of a power packet dispatching system, which has been proposed by authors' group with being apart from the conventional power distribution system. Here is focused on the function to establish the security of power packet dispatching for prohibiting not only information but also power of power packet from being stolen by attackers. For the purpose of protecting power packets, we introduce a simple encryption of power packets before sending them. Encryption scheme based on chaotic signal is one possibility for this purpose. This paper adopts the Differential Chaos Shift Keying (DCSK) scheme for the encryption, those are partial power packet encryption and whole power packet encryption.
\end{abstract}

\maketitle 
\section{Introduction}
\label{sec:intro}
\begin{figure}[htb]\centering   
\includegraphics[width=90mm]{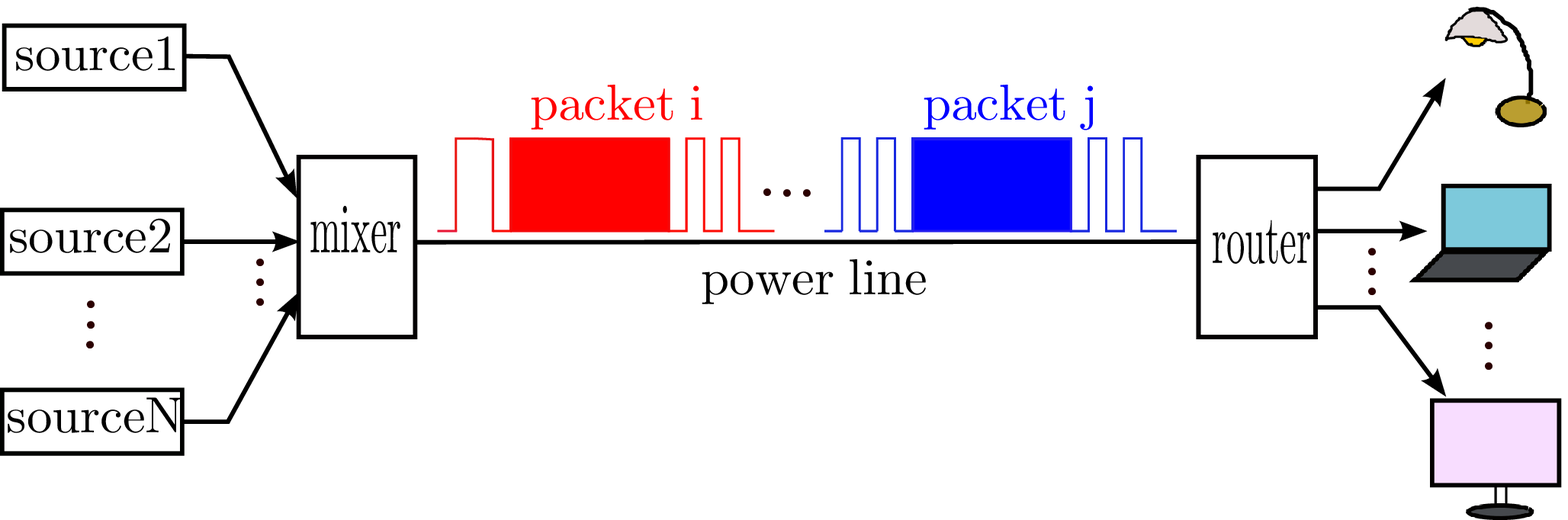}
\caption{Basic configuration of power packet dispatching system. Power packets are generated in the mixer and transmitted to the router. The power contained in each power packet is dispatched to the load designated by the information tags of the power packet.}
\label{fig:ppds}
\end{figure}

\begin{figure}[htb]\centering   
\includegraphics[width=90mm]{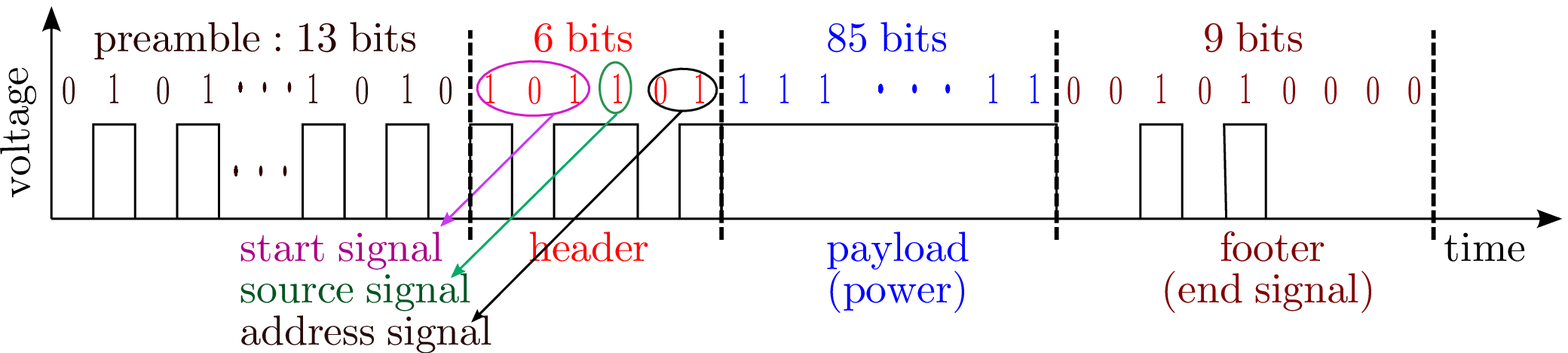}
\caption{Basic composition of one power packet. The preamble is several cycles of mixer clock signal used to achieve the clock synchronization, the header includes useful information, the payload carries power and the footer indicates the end of one power packet.}
\label{fig:powerpacket}
\end{figure}

In the context of the growth of renewable power sources in electricity generation, a novel power packet dispatching system has been proposed for the purpose of managing low power renewable power sources together with commercial power sources and supplying power based on demands ~\cite{hikihara,takuno2010,ta2011,takahashi}. The basic configuration of the power packet dispatching system, which consists of DC power sources, a mixer, power line, a router, and loads, is illustrated in Fig.~\ref{fig:ppds}. The mixer produces power packets by switching selected power sources on and off based on its clock frequency. The proposed configuration of one power packet is presented in Fig.~\ref{fig:powerpacket}, which includes information tags and payload \cite{zhouCTA}. In particular, useful information, such as the power source and power destination (load address), are the main part of the header. The footer is the end signal of one power packet and the payload carries electric power. Power packets are transmitted from a mixer to a router through a power line in the Time-Division Multiplexing (TDM) fashion. The router is in charge of determining the route for each power packet according to the information attached in the header and transferring power to the requested load \cite{takuno2010,Tashiro,takarouter}. For the sake of recognition of the packet signal in the router so as to determine the route for power packets precisely, the clock synchronization between the mixer and router is required. Different clock synchronization methods have been employed in the power packet dispatching system. In \cite{Tashiro} the clock synchronization has been achieved by a clock recovery circuit and in \cite{fujii} an external clock line is adopted for external synchronization. However, the line connection for synchronization limits the extension of system distribution, because of the delay and noise generated by clock lines. The failure in synchronization caused by the deficiency of external clock line is explained theoretically in \cite{nawata}. In \cite{zhouletter}, a first order control of digital clock synchronization is applied and the clock synchronization is achieved automatically between the mixer and router with the aid of the preamble of power packet. The preamble contains several cycles of the mixer clock signal. The performance of clock synchronization is improved by a second order control scheme in \cite{zhouCTA}.
\\\indent The security of power packet dispatching is defined as protecting the content of power packet, which consists of both information tags and power, from unknown consumer (attacker). For the system in Fig.~\ref{fig:ppds}, when an attacker who exists between the mixer and the router receives power packets, he may deduce the packet signal and then obtain private information of consumers. With respect to power, the attacker can steal the power if he connects some devices directly to the power line. In other words, there is no security in the basic power packet dispatching system according to the definition of security. We can easily imagine that the attacker may even change the information purposely and then send the tampered packet to the router. In doing so, he may give rise to wrong power dispatching, which will possibly cause a damage to the running system. Considering the consumers' privacy and safety of the system, the security of power packet between the mixer and router should be established. For this purpose, we propose to encrypt power packets before sending them. 
\\\indent Spread spectrum communications have been widely used in secure communication field. The uniform spreading of spectrum is one of the desired features in encryption. This is because the spectrum may be one of the clues for decryption. Besides that, a message may be hidden in the background noise by spreading its bandwidth with coding and by transmitting signal at a low averaged power, which is called a low-probability-of-intercept~\cite{proakis}. Chaotic sequences are very appealing for the use as carriers in Direct-Sequence Spread Spectrum (DSSS) system due to their broad band spectrum. They can spread spectrum more uniformly compared with the conventional sequences such as the Gold sequence and Walsh-Hadamard sequence as described in \cite{takahashiumeno}. Power spread is also considered in \cite{hamill,alex}. The possibility of synchronization between chaotic systems was reported by Pecora and Carroll \cite{pecora,pecoratwo,pecorathree}. After that, many chaos-based communication systems had been proposed and analyzed \cite{takahashiumeno,parlitz,kevin,dedieu,kohda,kolumban1996differential,tao,kolumban2,tao2,chen,stavrinides,ryo,kaddoum2,dcskbook,kaddoum}. In the power packet dispatching system, power packet is generated in digital form. Hence, digital encryption scheme is required. Chaos Shift Keying (CSK) scheme with coherent detection was firstly proposed in \cite {parlitz,dedieu} to encode digital symbols with chaotic signals. However, the sensitive dependence of chaotic signals upon initial condition makes it very difficult to replica signal in the receiver \cite{dcskbook}. Therefore, in chaos-based communication systems, non-coherent detection of received signal has advantage over coherent detection. The Differential Chaos Shift Keying (DCSK) scheme is a typical non-coherent chaos-based communication scheme \cite{kolumban1996differential}, which also solves the problem of threshold shift in non-coherent CSK system \cite{dcskbook}. In view of features of the DCSK scheme, we investigate the power packet encryption using DCSK scheme to estabish the security of power packet dispatching in this paper.
\\\indent The rest of the paper is organized as follows. In Section 2, the operating principles of DCSK scheme are summarized according to references 
\cite{kolumban1996differential,dcskbook} for its application to power packet encryption. Section 3 presents two power packet encryption methods using DCSK scheme, which are partial power packet encryption and whole power packet encryption. At the same time, the security status of power packet is also discussed. Power transfer through the modulator is analyzed and then examined through simulations in Section 4. The final section is devoted for the conclusions.
\section{Differential Chaos Shift Keying}
\label{sec:DCSK}
\begin{figure}[htb]
\begin{center}
\subfigure[ ]{
\label{fig:DCSKmod}
\includegraphics[width=5.5cm]{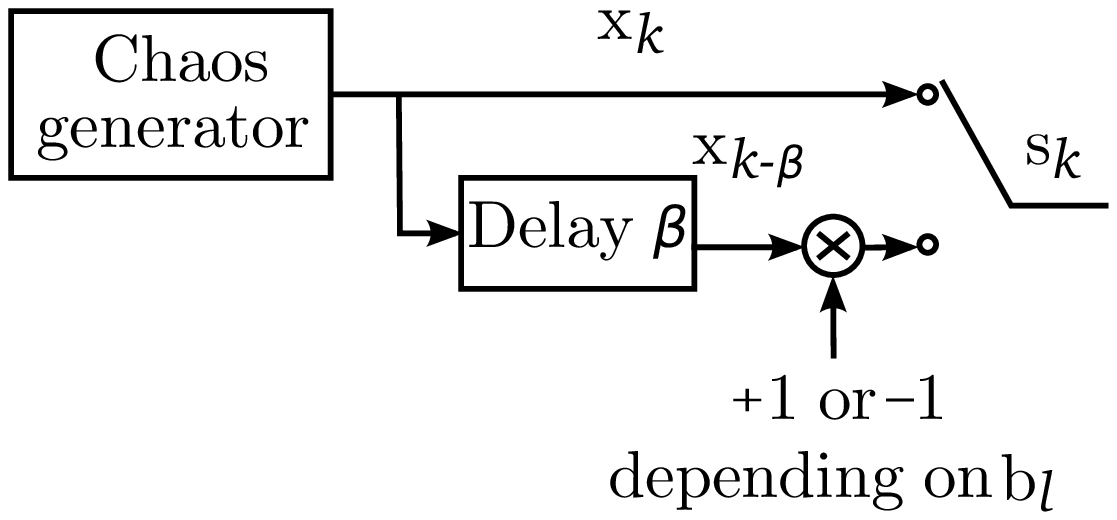}}
\subfigure[ ]{
\label{fig:DCSKdemod}
\includegraphics[width=6.5cm]{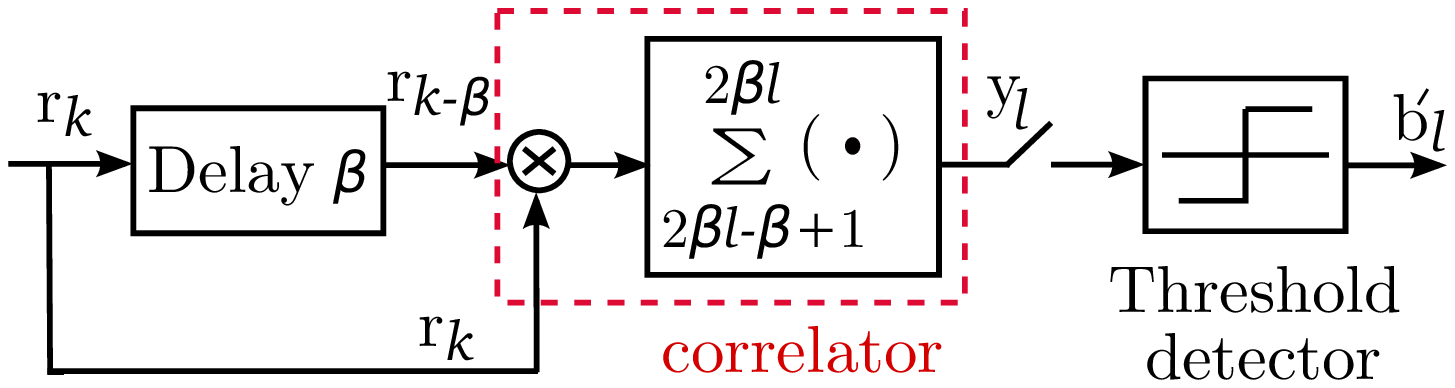}}
\end{center}
 \caption{Block diagram of a non-coherent single-user DCSK system. (a) modulator; $x_{k}$ and $x_{k-\beta}$ are chaotic signal and its half-bit ($\beta$) delayed version, respectively. They are used to represent the information bit $b_{l}$. $s_{k}$ is the modulator output. (b) demodulator. $r_{k}$ and $r_{k-\beta}$ are received signal and its half-bit delayed version, respectively. $y_{l}$ is the convolution of $r_{k}$ and $r_{k-\beta}$ during the $l$-th bit period while $b_{l}'$ is the recovered $l$-th information bit. $2\beta$ is the spreading factor in the DCSK scheme.}
\end{figure}

\begin{figure}[htb]\centering   
\includegraphics[width=70mm]{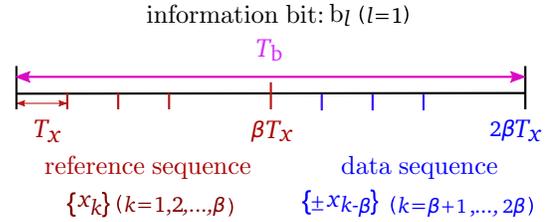}
\caption{Representation of one information bit $b_{1}$ using DCSK scheme. The information bit period $T_{\rm b}$ is divided into two equal time slots. $b_{1}$ is represented by two consecutive chaotic sequence: reference sequence and data sequence ($\{x_{k}\}$ and $\{\pm x_{k-\beta}\}$). The sign of data sequence is determined by $b_{1}$ (if $b_{1}$= `1', the data sequence is $\{x_{k-\beta}\}$, otherwise, it is $\{-x_{k-\beta}\}$). $T_{x}$ indicates the time interval between two adjacent points of chaotic sequence ($T_{\rm b}$=$2\beta$$T_{x}$). }
\label{fig:dcskonebit}
\end{figure}

The operating principles of the Differential Chaos Shift Keying (DCSK) scheme was reported in \cite{kolumban1996differential,dcskbook}. Figure~\ref{fig:DCSKmod} presents the block diagram of a DCSK modulator. In the scheme, every information bit period is divided into two equal time slots, so that every bit $b_{l}$ can be represented by two consecutive chaotic sequences: reference sequence and data sequence. Here, $l\in \mathbb{N_{\rm 1}}$ denotes the serial number of information bits. In the first half bit period, the chaotic signal $x_{k}$ is transmitted ($k\in \mathbb{N_{\rm 1}}$ indicates the serial number of sample points of the chaotic signal). During the second half bit period the data sequence is transmitted. The data sequence is determined by $b_{l}$  and the reference sequence. When the information bit to be sent is `1', the data sequence is identical to the reference sequence. Whereas, if the information bit is `--1', the data sequence is inverted to the reference sequence. In this way, the data sequence carries information bits. The representation of one information bit $b_{1}$ using DCSK scheme is exemplified in Fig.~\ref{fig:dcskonebit}. Let $T_{\rm b}$ and $T_{x}$ represent the information bit period and the time interval between two adjacent points of chaotic signal, respectively. The relationship of $T_{\rm b}$ and $T_{x}$ is given as 
\begin{equation}
T_{\rm b}=2\beta T_{x},
\label{eq:relationship}
\end{equation}
where $2\beta$ is referred to as the spreading factor in the DCSK scheme ($\beta\in \mathbb{N_{\rm 1}}$). Consequenctly, the data sequence can be represented as $\{\pm x_{k-\beta}\}$, $k=\beta+1$, $\beta+2$, $...$ , $2\beta$ for $l=1$. The output signal of the DCSK modulator $s_{k}$, i.e., the transmitted signal, during the $l$-th bit period can be given as follows for $b_{l}=$`1',
\begin{equation}
s_{k}=
\begin{array}{r@{\,}l}
\mathrm{ } &\left\{
 \begin{array}{c@{\,}l}
  x_{k},\thinspace (l-1)2\beta+1\le k \le (l-1)2\beta+\beta, \\ \\ 
  x_{k-\beta},\thinspace (l-1)2\beta+\beta + 1\le k \le (l-1)2\beta+2\beta.
 \end{array}\right.
\end{array}
\label{eq:transsigpos}
\end{equation}
When a `--1' is sent, $s_{k}$ can be expressed in a likewise fashion, 
\begin{equation}
s_{k}=
\begin{array}{r@{\,}l}
\mathrm{ } &\left\{
 \begin{array}{c@{\,}l}
  x_{k},\thinspace (l-1)2\beta+1\le k \le (l-1)2\beta+\beta, \\ \\ 
  -x_{k-\beta},\thinspace (l-1)2\beta+\beta + 1\le k \le (l-1)2\beta+2\beta.
 \end{array}\right.
\end{array}
\label{eq:transsigneg}
\end{equation}

The block diagram of a DCSK demodulator \cite{dcskbook} is presented in Fig.~\ref{fig:DCSKdemod}. The output of the correlator $y_{l}$ can be obtained at the end of the $l$-th bit period by the following equation.
\begin{equation}
y_{l}=\sum\limits_{k=(l-1)2\beta+\beta+1}^{(l-1)2\beta+2\beta} r_{k}r_{k-\beta}.
\label{eq:correlator}
\end{equation}
Comparing $y_{l}$ with the threshold value of the threshold detector, the information bit can be recovered at the output of the demodulator as $b_{l}'$. Suppose that the transmitted signal $s_{k}$ passes through an Additive White Gaussian Noise (AWGN) channel \cite{dcskbook}. The received signal $r_{k}$ can be given as
\begin{equation}
r_{k}=s_{k}+\xi_{k},
\label{eq:rk}
\end{equation}
where $\xi_{k}$ represents the $k$-th sampling point of the white Guassian noise which is assumed to be of zero mean and variance as $N_{0}/2$. If a `+1' is transmitted, $s_{k}$ is as in Eq.~(\ref{eq:transsigpos}). Substituting Eqs.~(\ref{eq:transsigpos}) and~(\ref{eq:rk}) into Eq.~(\ref{eq:correlator}), $y_{l}$ can be described in more detail as follows. 
\begin{equation}
\begin{split}
y_{l}&=\sum\limits_{k=(l-1)2\beta+\beta+1}^{(l-1)2\beta+2\beta} r_{k}r_{k-\beta}\\
&=\sum\limits_{k=(l-1)2\beta+1}^{(l-1)2\beta+\beta} r_{k+\beta}r_{k}\\
&=\sum\limits_{k=(l-1)2\beta+1}^{(l-1)2\beta+\beta} (s_{k+\beta}+\xi_{k+\beta})(s_{k}+\xi_{k})\\
&=\sum\limits_{k=(l-1)2\beta+1}^{(l-1)2\beta+\beta} (x_{k}+\xi_{k+\beta})(x_{k}+\xi_{k})\\
&=\sum\limits_{k=(l-1)2\beta+1}^{(l-1)2\beta+\beta} x_{k}^2+\sum\limits_{k=(l-1)2\beta+1}^{(l-1)2\beta+\beta} x_{k}(\xi_{k}\\
& \quad +\xi_{k+\beta})+\sum\limits_{k=(l-1)2\beta+1}^{(l-1)2\beta+\beta} \xi_{k}\xi_{k+\beta}.\\
\end{split}
\label{eq:correlator2}
\end{equation}
Let $\sum\limits_{k=(l-1)2\beta+1}^{(l-1)2\beta+\beta} x_{k}^2$, $\sum\limits_{k=(l-1)2\beta+1}^{(l-1)2\beta+\beta} x_{k}(\xi_{k}+\xi_{k+\beta})$, and $\sum\limits_{k=(l-1)2\beta+1}^{(l-1)2\beta+\beta} \xi_{k}\xi_{k+\beta}$ be denoted by $A$, $B$, and $C$, respectively. $A$ is related only to the chaotic signal $x_{k}$ and is positive. $B$ and $C$ are noise interference. However, if a `--1' is sent, $s_{k}$ is as in Eq.~(\ref{eq:transsigneg}). Accordingly, $y_{l}$ becomes 
\begin{equation}
\begin{split}
y_{l}&=\sum\limits_{k=(l-1)2\beta+1}^{(l-1)2\beta+\beta} (-x_{k}^2)+\sum\limits_{k=(l-1)2\beta+1}^{(l-1)2\beta+\beta} [-x_{k}(\xi_{k}+\xi_{k+\beta})]\\
& \quad +\sum\limits_{k=(l-1)2\beta+1}^{(l-1)2\beta+\beta} \xi_{k}\xi_{k+\beta}.\\
\end{split}
\label{eq:correlatorneg}
\end{equation}
Let $\sum\limits_{k=(l-1)2\beta+1}^{(l-1)2\beta+\beta} (-x_{k}^2)$ and $\sum\limits_{k=(l-1)2\beta+1}^{(l-1)2\beta+\beta} [-x_{k}(\xi_{k}+\xi_{k+\beta})]$ be denoted by $A'$ and $B'$, respectively. $A'$ is also related only to the chaotic signal $x_{k}$ but it is negative. Since AWGN channel is assumed between the modulator and demodulator, it can be deduced that both $B$ (or $B'$) and $C$ have a zero mean. Therefore, the threshold value of the detector can be set at zero optimally. As a consequence, the information bit can be recovered through the detector as
\begin{equation}
b_{l}'=
\begin{array}{r@{\,}l}
\mathrm{ } &\left\{
 \begin{array}{c@{\,}l}
 `+1\textrm',\qquad  y_{l} > 0, \\ \\ 
 `-1\textrm',\qquad  y_{l} < 0.
 \end{array}\right.
\end{array}
\label{eq:recovsig}
\end{equation}

\section{Power Packet Encryption}
Given the principles of DCSK scheme in section~\ref{sec:DCSK}, we apply the scheme to encrypt power packet aiming to establish the security of power packet dispatching. When the scheme is applied to power packet encryption, the packet signals are encrypted as information bits in a conventional DCSK modulator. In this section, two encryption methods for power packet will be discussed, i.e., partial power packet encryption and whole power packet encryption. From the principle of DCSK demodulator, the spreading factor ($2\beta$) should be obtained by the demodulator in advance so that the information bits can be recovered correctly. We refer to $2\beta$ as key in our encryption methods. 
\subsection{Partial Power Packet Encryption}
\label{sec:part}
\begin{figure}[tb]\centering   
\includegraphics[width=90mm]{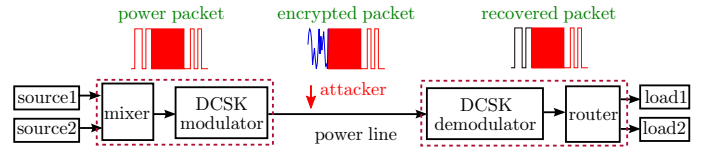}
\caption{Diagram of system with partial power packet encryption. The power packet is partially encrypted by the DCSK modulator. The encrypted packet is recovered through the demodulator before it arrives to the router. Finally, the power contained in the power packet is delivered to the desired load.}
\label{fig:moddemodsys}
\end{figure}

In the power packet dispatching system, the synchronization is essential to handshake between the sender and receiver of power packet, that is, the mixer and router. As shown in Fig.~\ref{fig:powerpacket}, the preamble is designed for achieving the clock synchronization between the mixer and router \cite{zhouletter}. At the meantime, from the viewpoint of privacy protect, the header should be hidden during transmission, because the header may include consumers' private information. Therefore the security of information is possibly established, provided that the preamble and header are protected. In partial power packet encryption, only preamble and header are encrypted using DCSK scheme. The rest of the packet (payload and footer) remains unencrypted. The encryption method is illustrated in Fig.~\ref{fig:moddemodsys}. The mixer is integrated with a DCSK modulator and the router is integrated with a DCSK demodulator. As shown in this figure, the power packet generated by the mixer is partially encrypted through the DCSK modulator. The encrypted packet is then transmitted. In demodulator, the encrypted packet can be recovered if the key is obtained beforehand as mentioned in section~\ref{sec:DCSK}. The whole packet can be recovered, not only because the recovered preamble and header suffice to determine the route of power packet in the router, but also based on the fact that the unencryption of the payload and footer guarantees the power in power packet. 
\\\indent Under partial encryption, assume that an attacker exists in the system during power packet transmission. The security status of information and power can be analyzed. Here, we focus on the information contained in the preamble and header without considering the footer, since the footer is the end signal of a power packet. When the packet is caught by the attacker, he can not obtain information because of the disability of recovering the packet signals without the key. Therefore, the security of information is achieved. Nevertheless, the power can be stolen by the attacker if he connect some device directly to the power line. It means the security of power can not be achieved in this method. In addition, the power of power packet is carried by the payload as mentioned in section~\ref{sec:intro}. Therefore, whole power of packet may be stolen by the attacker since the payload is transmitted without encryption.
\subsection{Whole Power Packet Encryption}
\label{sec:whole}
\begin{figure}[tb]\centering   
\includegraphics[width=90mm]{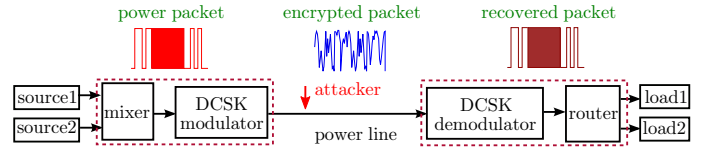}
\caption{Diagram of system with whole power packet encryption. The whole power packet is encrypted by the DCSK modulator. The encrypted packet is recovered through the demodulator before it arrives to the router. Finally, the power contained in the power packet is delivered to the desired load.}
\label{fig:wholemoddemodsys}
\end{figure}

\begin{figure}[tb]\centering   
\includegraphics[width=60mm]{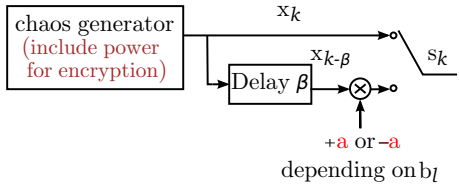}
\caption{Block diagram of modulator used in whole power packet encryption. This is the modified version of the original DCSK modulator. $a$ is the absolute voltage amplitude of the packet signal $b_{l}$.}
\label{fig:wholemodDCSK}
\end{figure}

In order to further improve the security of power packet dispatching, we propose to encrypt the whole power packet.  Figure~\ref{fig:wholemoddemodsys} shows the whole power packet encryption method. It is clear that there is no difference in the security of information between these two methods, since the preamble and header are encrypted in both methods. Next, we will discuss about the power. The power can still be stolen by the attacker if he connects some devices directly to the power line. However, we should bear in mind that payload carries the power of packet. Thus what calls for special attention is that once the payload is encrypted, the power of encrypted packet may be changed.
\\\indent The whole encryption method encrypts both information tags and payload. It is different from the original DCSK scheme, where the tranferred signal is digital data without physical quantity. In order to transfer power, the DCSK modulator is modified as in Fig.~\ref{fig:wholemodDCSK}. In the chaos generator, we suppose the generated chaotic signal carries power for encryption. The data sequence becomes the product of chaotic signal and the packet signal. The voltage amplitude of packet signal is `+a' or `--a' in volts. Let the payload consists of $N_{\rm b}$ bits packet signal. Then the average output power of modulator $P_{\rm modout}$ in watts (W) can be calculated during payload according to Eq.~(\ref{eq:transsigpos}) with replacing $x_{k-\beta}$ by $ax_{k-\beta}$ as follows:
\begin{equation}
\begin{split}
P_{\rm modout}&=\frac{\sum\limits_{l=1}^{N_{\rm b}} (\sum\limits_{m=1}^{\beta} x_{lm}^{2}T_{x}+\sum\limits_{m=1}^{\beta} a^{2}x_{lm}^{2}T_{x})}{N_{\rm b}T_{b}}\\
& \thickspace =\frac{(1+a^{2})\sum\limits_{l=1}^{N_{\rm b}} \sum\limits_{m=1}^{\beta} x_{lm}^{2}}{2\beta N_{\rm b}},
\end{split}
\label{eq:modout average power}
\end{equation}
where $x_{lm}$ denotes the $m$-th sample point of chaotic signal in the first half bit period of the $l$-th bit packet signal. Equation~(\ref{eq:modout average power}) shows the possibility of rescaling the  output power of modulator $P_{\rm modout}$ by the voltage amplitude of packet signal $a$, the spreading factor $2\beta$, the bit number of packet signal during payload $N_{\rm b}$ and chaotic signal. From the viewpoint of the amount of stolen power, we say that the power of power packet under whole encryption becomes more secure compared to the partial encryption method.
\\\indent We consider the original power packet is the input of the modulator and thus the average input power of the modulator during payload is $a^{2}$ in watts, since the voltage of packet signal is constant at $a$ V throughout the payload. Power efficiency of the modulator $\eta_{\rm mod}$ is definded as below,
\begin{equation}
\eta_{\rm mod}=\frac{(1+a^{2})\sum\limits_{l=1}^{N_{\rm b}}\sum\limits_{m=1}^{\beta} x_{lm}^{2}}{2\beta N_{\rm b}a^{2}}.
\label{eq:power ratio}
\end{equation}
It corresponds to the ratio of the output power divided by the input power of the modulator. $\eta_{\rm mod}$ is possible to rescale by $a$, $2\beta$, $N_{\rm b}$, and $x_{k}$. As a result, the range of input power can be expanded to meet the demand of power from loads. This might be another benifit of whole packet encryption using DCSK scheme.
\section{Simulation Results of Transferred Power}
\begin{figure}[htb]\centering   
\includegraphics[width=70mm]{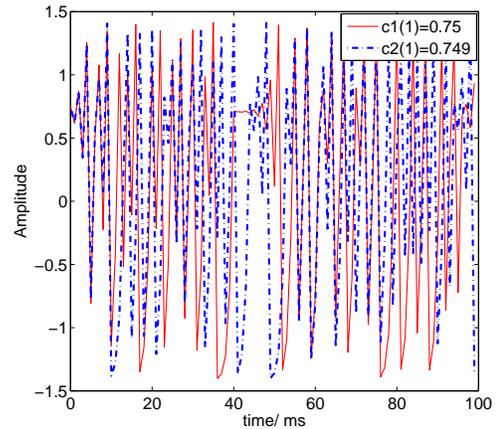}
\caption{Waveforms of chaotic signals generated by iterative map $x_{k+1}=\sqrt{2}(1-x_{k}^2)$ with different initial values. c1(1)=0.75, c2(1)=0.749. The time interval between signals is $T_{\rm c1}=T_{\rm c2}=0.001$ s and the simulation time is $0.1$ s, which implies the length of signals plotted here is $N_{\rm c1}=N_{\rm c2}=100$.} 
\label{fig:diffini}
\end{figure}

\begin{figure}[htb]
\centering
\includegraphics[width=7cm]{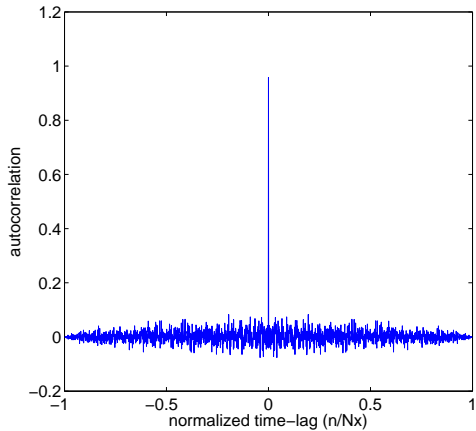}
\caption{Normalized approximated auto-correlation of $x_{k}$, which is generated by iterative map $x_{k+1}=\sqrt{2}(1-x_{k}^2)$ with $x_{1}=0.75$ and the time step is $T_{x}=0.001$ s. The generated samples are of finite length 1000.}
\label{fig:autocorr}
\end{figure}

\begin{figure}[htb]\centering   
\includegraphics[width=7cm]{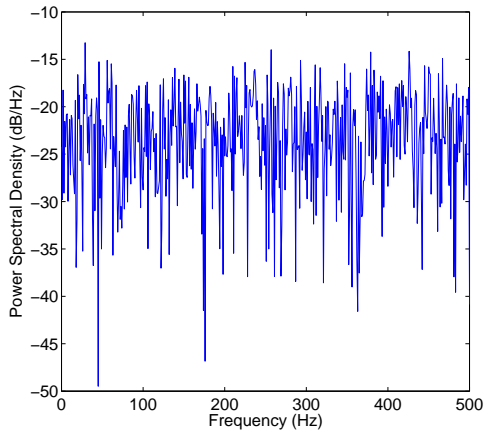}
\caption{Approximated power spectral density of $x_{k}$, which is generated by the iterative map $x_{k+1}=\sqrt{2}(1-x_{k}^2)$ with $x_{1}=0.75$ and time step is $T_{x}=0.001$ s. The generated samples are of finite length 1000.}
\label{fig:powerspeccsig}
\end{figure}

\begin{figure*}[htb]
\begin{center}
\subfigure[ ]{
\includegraphics[width=7.5cm]{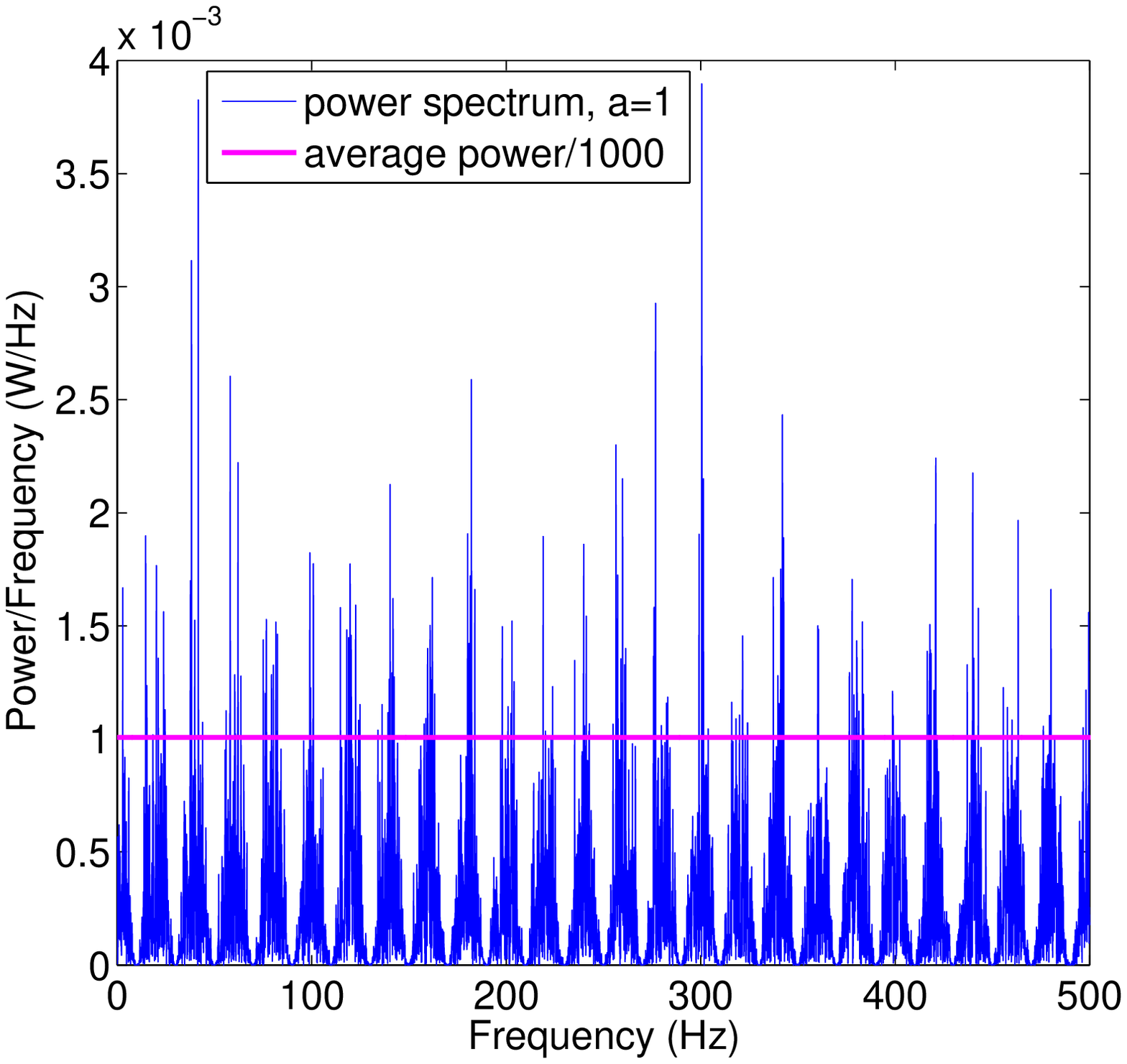}}
\subfigure[ ]{
\includegraphics[width=7.5cm]{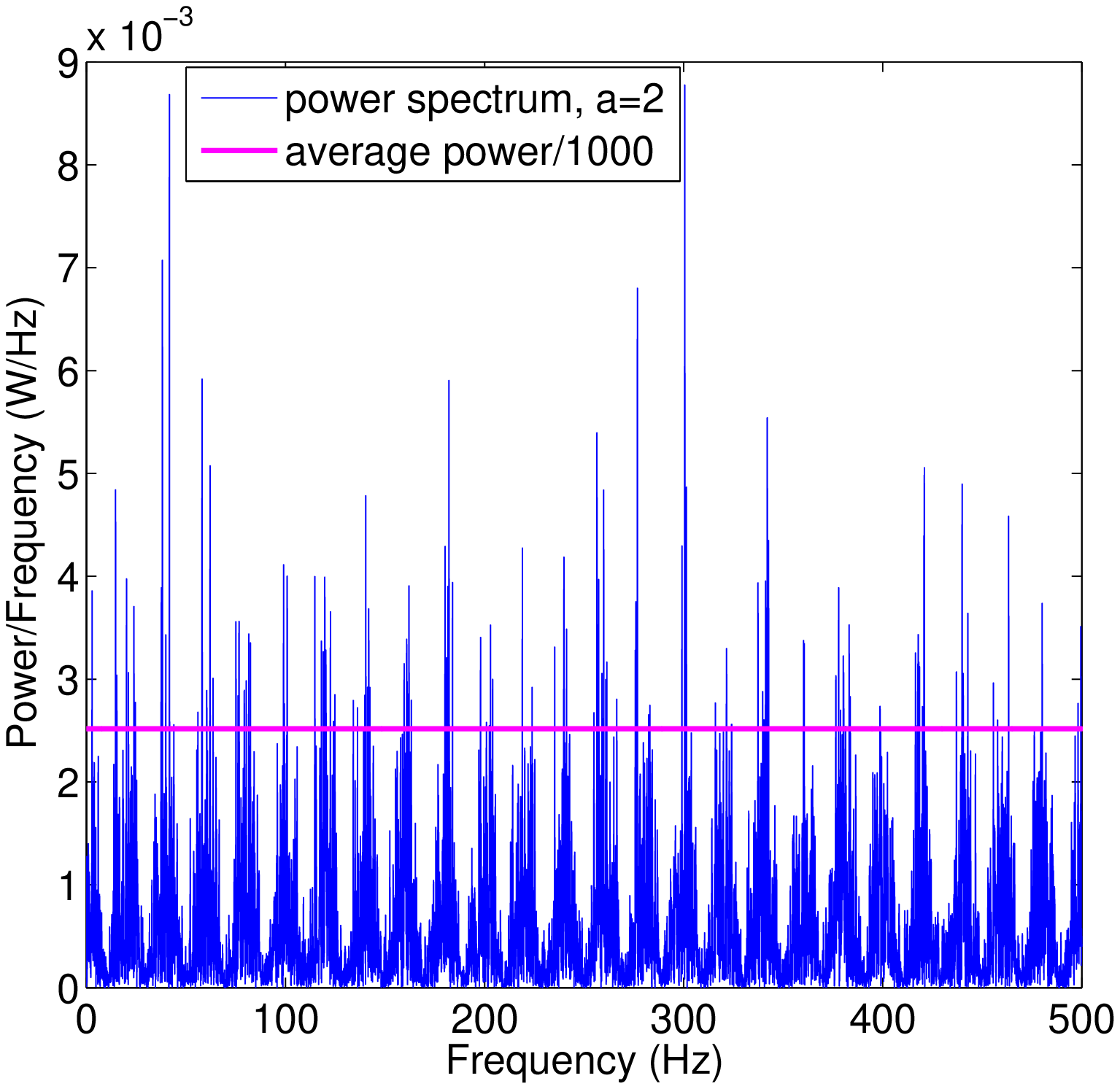}}
\subfigure[ ]{
\includegraphics[width=7.5cm]{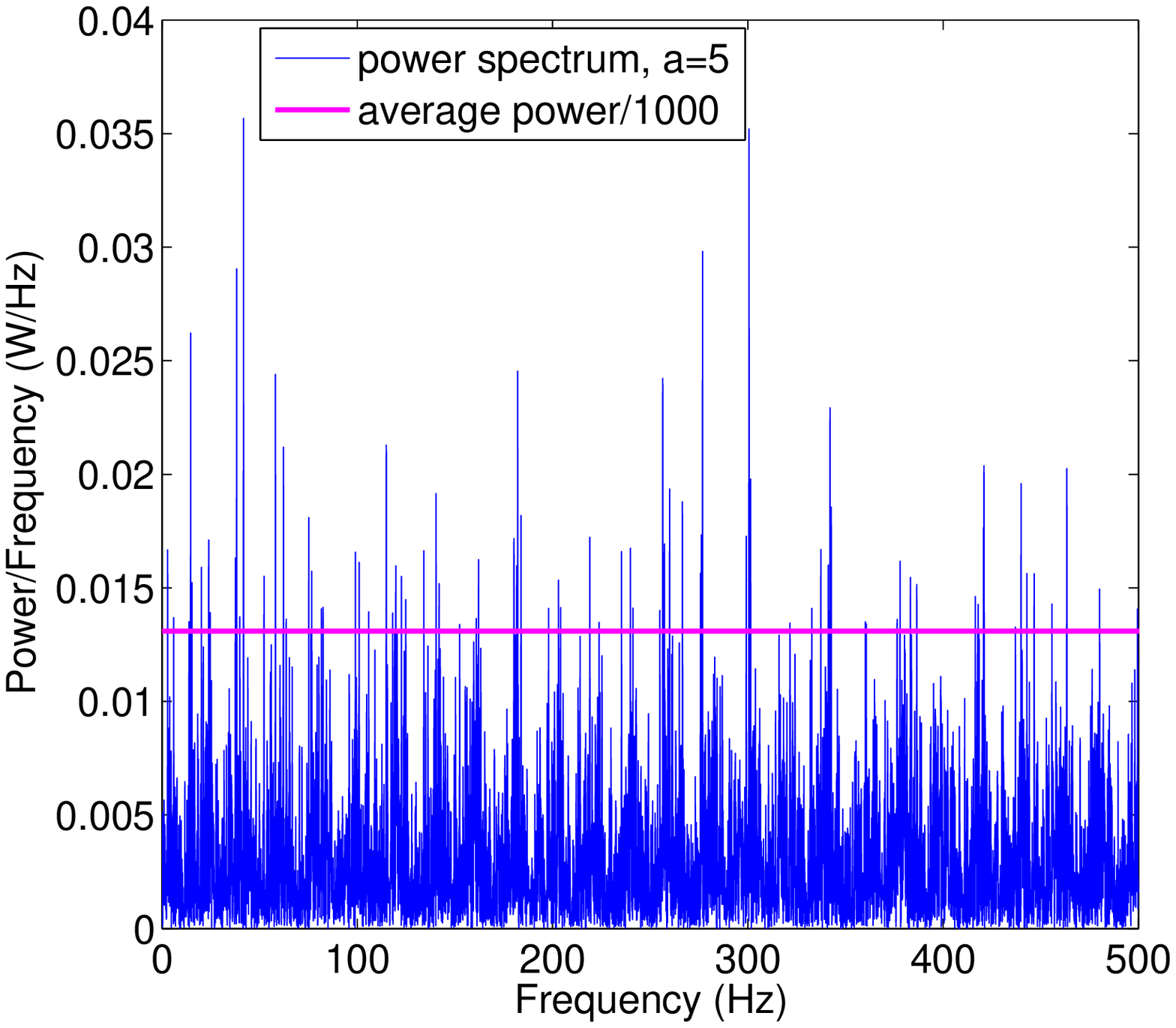}}
\subfigure[ ]{
\includegraphics[width=7.5cm]{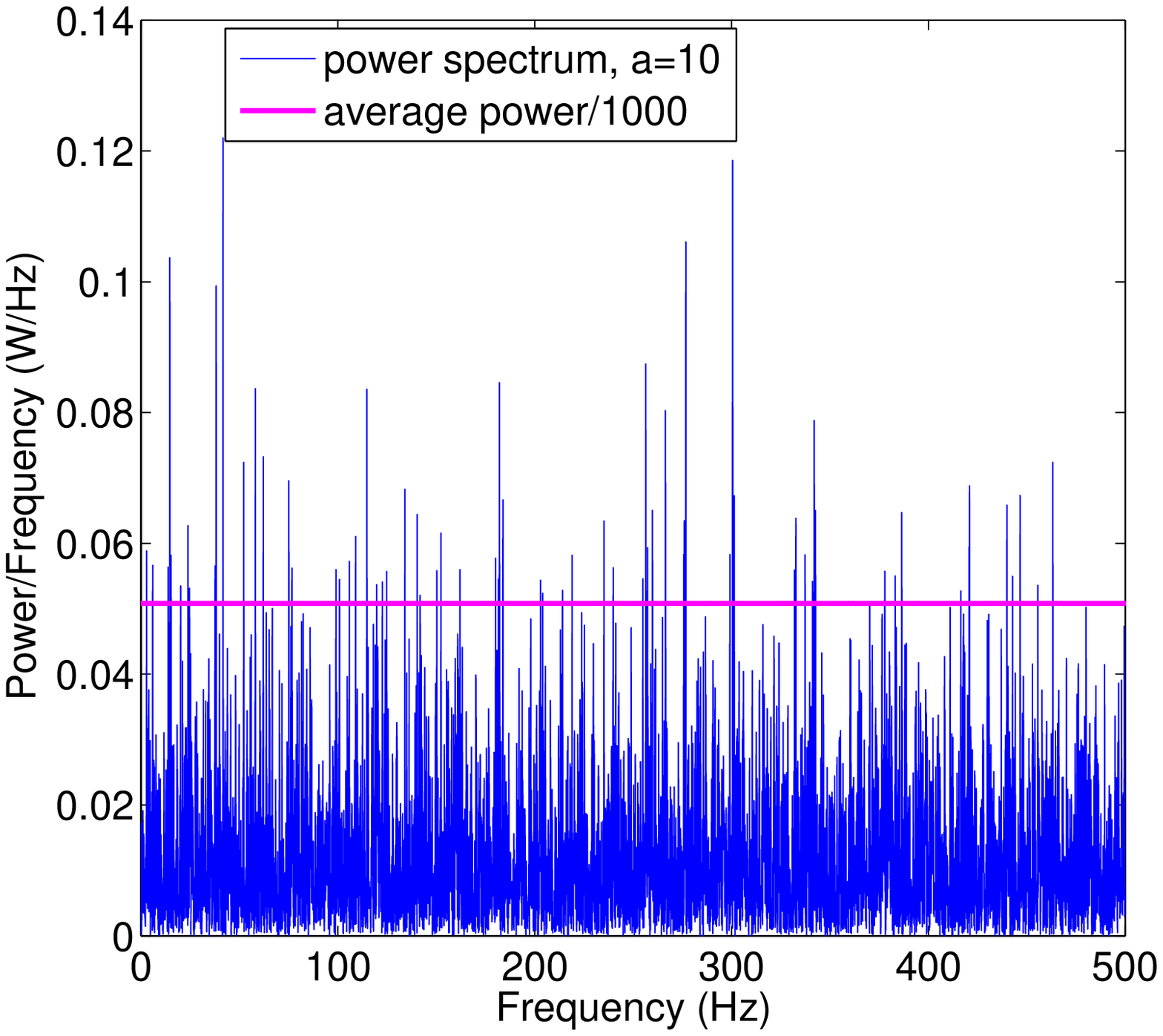}}
\end{center}
 \caption{Power spectrum of DCSK modulator output signal $s_{k}$ in simulations with different values of $a$, while $T_{\rm sam}=0.001$ s and $2\beta=100$. The data is limited in payload duration. $T_{\rm sam}$ is the sampling period in simulations, $2\beta$ the spreading factor and $a$ the absolute amplitude of packet signal in volts. (a) $a=1$; (b) $a=2$; (c) $a=5$; (d) $a=10$. The magenta line represents the value of average power of $s_{k}$ in payload duration.}
\label{fig:a}
\end{figure*}

\begin{figure*}[htb]
\begin{center}
\subfigure[ ]{
\includegraphics[width=7.5cm]{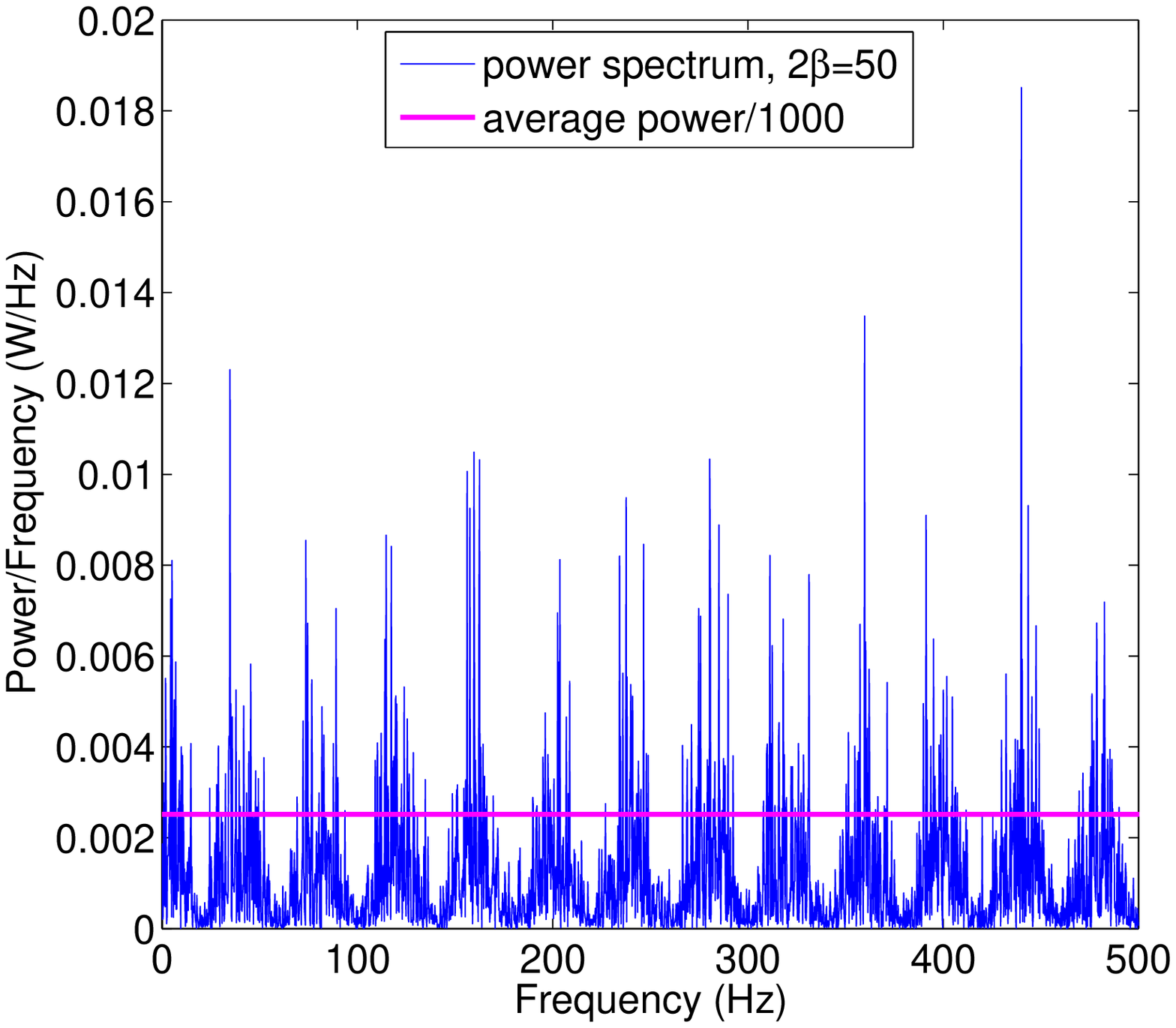}}
\subfigure[ ]{
\includegraphics[width=7.5cm]{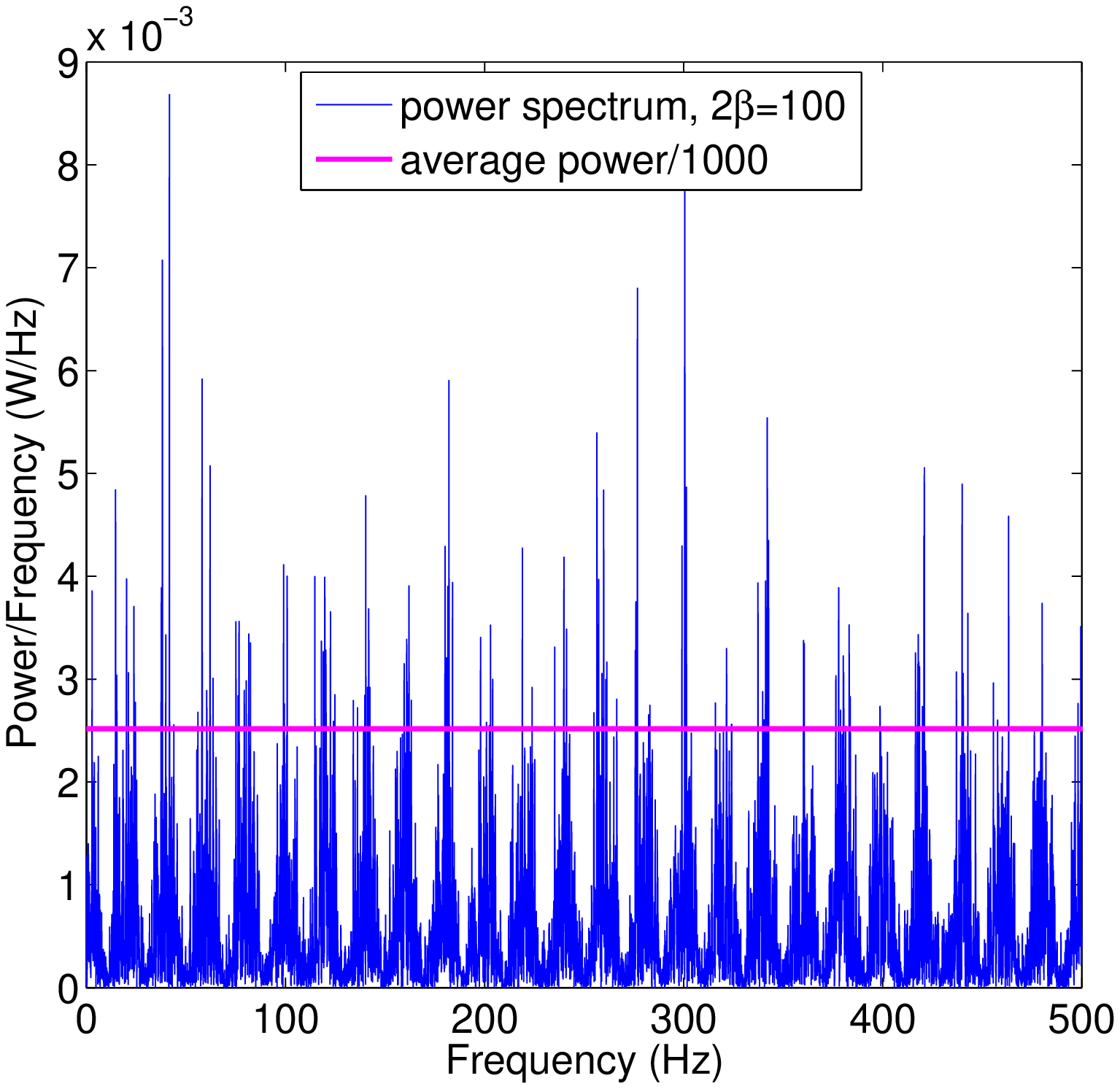}}
\subfigure[ ]{
\includegraphics[width=7.5cm]{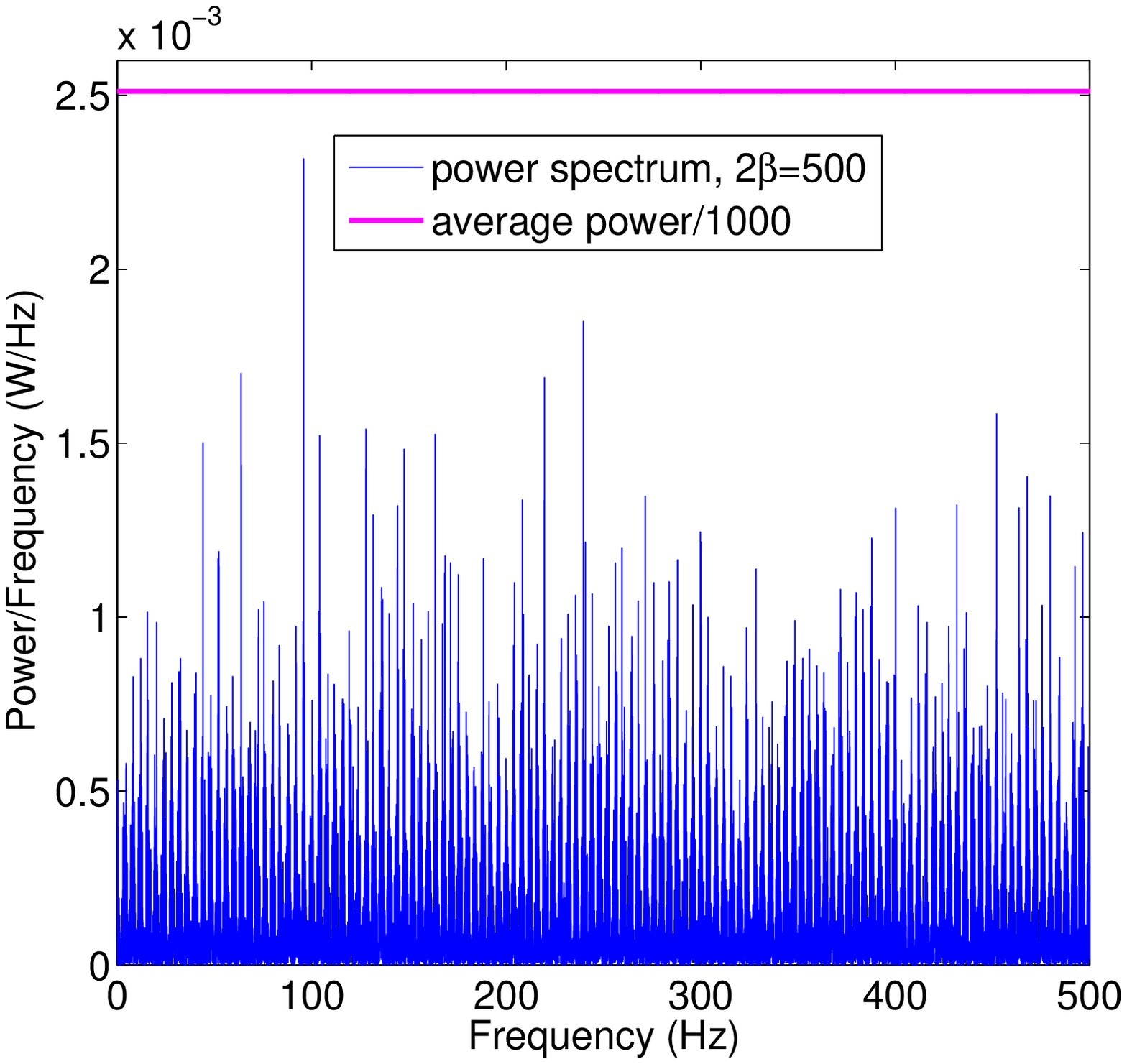}}
\subfigure[ ]{
\includegraphics[width=7.5cm]{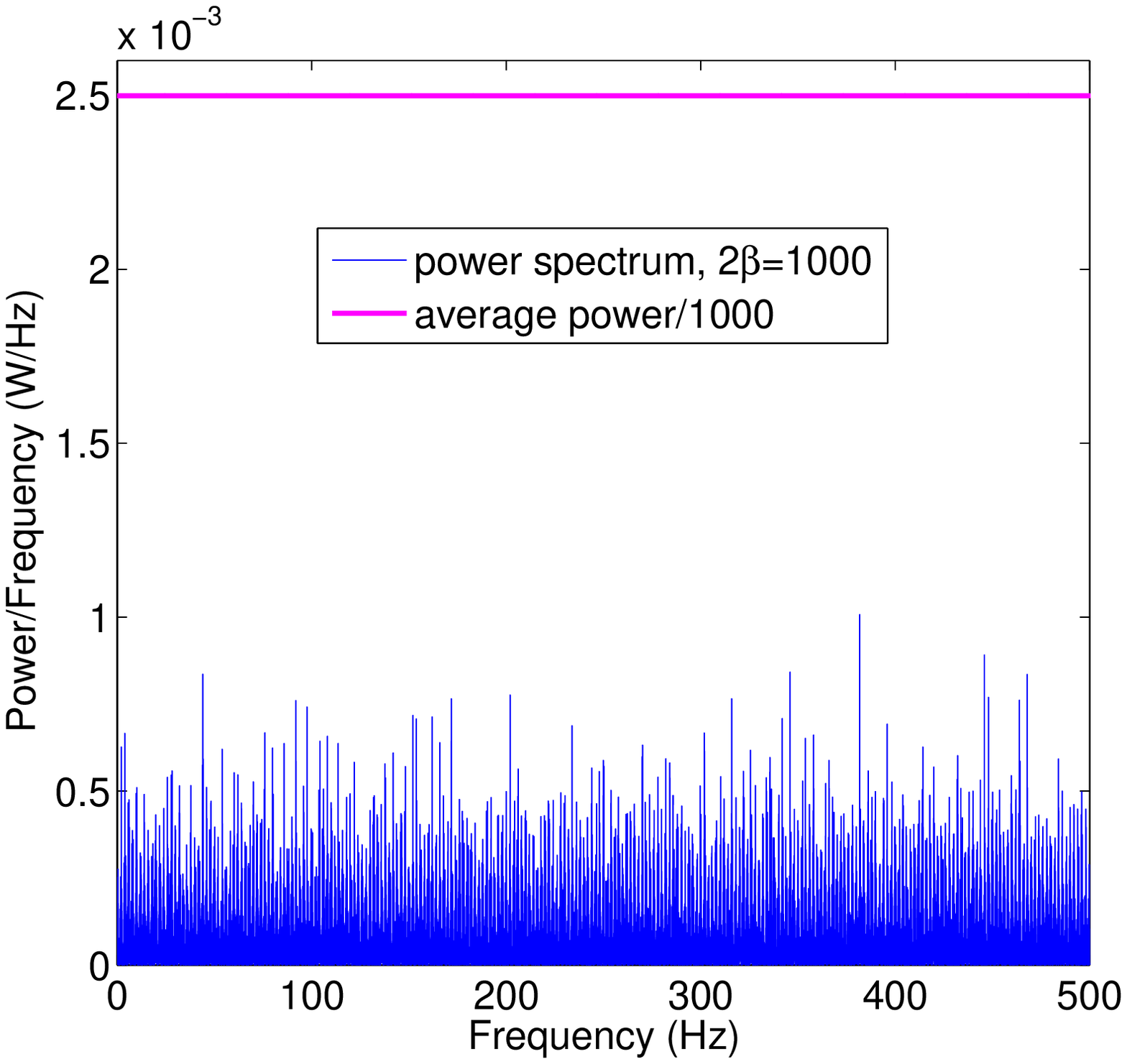}}
\end{center}
 \caption{Power spectrum of DCSK modulator output $s_{k}$ in simulation with different values of $2\beta$, while $T_{\rm sam}=0.001$ s and $a=2$. The data is limited in payload duration. $T_{\rm sam}$ is the sampling period in simulations, $2\beta$ the spreading factor and $a$ the absolute amplitude of packet signal in volts. (a) $2\beta=50$; (b) $2\beta=100$; (c) $2\beta=500$; (d) $2\beta=1000$.}
\label{fig:2beta}
\end{figure*}

In order to confirm the power transferred through DCSK modulator in whole encryption case, we build a model of DCSK modulator in Simulink. In simulations, the power packet shown in Fig.~\ref{fig:powerpacket} is adopted. From Eq.~(\ref{eq:modout average power}), the power spectrum of $s_{k}$ is related to the chaotic signal. Hence we begin with introduction of the chaotic signal utilized in simulations for analyzing the average power of $s_{k}$ in payload duration.
\\\indent In simulations,  the chaotic signal $x_{k}$ is produced by the following normalized improved logistic map \cite{dcskbook}, 
\begin{equation}
x_{k+1}=\sqrt{2}(1-x_{k}^2).
\label{eq:logistic}
\end{equation}
The sensitive dependence upon initial conditions of the produced chaotic signal is presented in Fig.~\ref{fig:diffini}. Both c1 and c2 in Fig.~\ref{fig:diffini} are produced by Eq.~(\ref{eq:logistic}) but with different initial conditions as c1(1)=0.75 and c2(1)=0.749. 
Next, the auto-correlation function of $x_{k}$ ($R_{xx}[n]$) is defined as
\begin{equation}
R_{xx}[n]=E(x_{k}x_{k-n}),
\label{eq:autox}
\end{equation}
where $n$ represents the time-lag and $E(\cdot)$ is the expected value operator. The power spectral density function is the Fourier transform of the auto-correlation function. Shown in Figs.~\ref{fig:autocorr} and~\ref{fig:powerspeccsig} are the numerical approximations of the auto-correlation function and power spectral density of $x_{k}$ using finite-length ($N_{x}=1000$) signals, respectively.    
Figure~\ref{fig:autocorr} indicates $x_{k}$ is almost random. The power spectral density in Fig.~\ref{fig:powerspeccsig} shows the wide-band property of $x_{k}$, which implies the aperiodicity.
\\\indent  As mentioned above, the power packet in \cite{zhouCTA} is employed, thus $N_{\rm b}=85$. We set the initial value of the chaotic signal $x_{1}$ at 0.75 without loss of generality. According to Eq.~(\ref{eq:logistic}), $x_{k}$ is limited in the range of $[-1.4142,+1.4142]$. In addition, if the bit length of packet signal $T_{\rm b}$ is determined, then the values of $x_{k}$ during payload can be obtained. The dependency of the output power of modulator $P_{\rm modout}$ on $a$ and $2\beta$ can be examined in simulations. We set $T_{x}=T_{\rm sam}=0.001$ s without loss of generality. $T_{\rm sam}$ represents the sampling period in simulations.
\\\indent When we fix $2\beta=100$, the power spectrums of $s_{k}$ with different values of $a$ ($a=1$, $a=2$, $a=5$, and $a=10$) are obtained as shown in Fig.~\ref{fig:a}.
The magenta lines in these figures show the average power of $s_{k}$, which means $P_{\rm modout}$. We refer to the value of $P_{\rm modout}$ obtained from the power spectrum of $s_{k}$ through simulations as $P_{\rm modoutsim}$, as given in Table~\ref{tab:Pspayb}. The mean values of $x_{k}$ and $x_{k}^{2}$ are $E[x_{k}]=0.0018$ and $E[x_{k}^{2}]=1.0067$ at $T_{\rm sam}=0.001$ s and $2\beta=100$. Then $P_{\rm modout}$ can be numerically calculated by Eq.~(\ref{eq:modout average power}) as in Table~\ref{tab:Pspayb}. In Table~\ref{tab:Pspayb}, with the increase of $a$, $P_{\rm modout}$ tends to be half of the input power $a^{2}$. This can be explained by substituing $E[x_{k}^{2}]=1.0067$ into Eq.~(\ref{eq:power ratio}).
\begin{table}[!h]
\caption{Power transferred through modulator with different values of $a$ while $2\beta$ is fixed at 100. $P_{\rm modoutsim}$ and $P_{\rm modout}$ corresponds to simulation result and calculation result, respectively.}
\vspace{3mm}
\centering
    \begin{tabular}{ | c | c | c | c | c |}
    \hline
   $a / V$ & 1 & 2 & 5 & 10\\ \hline
   $P_{\rm modoutsim} / W$& 1.007 & 2.517 & 13.09 & 50.84 \\ \hline
   $P_{\rm modout}/ W$& 1.0067 & 2.51675& 13.0871 &50.83835 \\ \hline
    \end{tabular}
\label{tab:Pspayb}
\end{table}
\\\indent Keeping $a=2$ the power spectrum of $s_{k}$ is considered for $2\beta$ ($2\beta=$50, 100, 500, and 1000). The power spectrums are obtained as in Fig.~\ref{fig:2beta}. Similarly, $P_{\rm modoutsim}$ are listed in Table~\ref{tab:p}. $E[x_{k}^{2}]$ is obtained through iteration as in Table~\ref{tab:Ex2}. $E[x_{k}^{2}]$ are different because the number of samples varies with $2\beta$. However, the difference is small due to the random feature of chaotic siganl. Then $P_{\rm modout}$ with different $2\beta$ can also be calculated according to Eq.~(\ref{eq:modout average power}), as given in Table~\ref{tab:p}. We can see that $P_{\rm modout}$ almost keep the same with the change of $2\beta$. It can be explained by Eq.~(\ref{eq:modout average power}) and Table~\ref{tab:Ex2}. According to Eq.~(\ref{eq:modout average power}), $P_{\rm modout}$ depends only on $E[x_{k}^{2}]$ when $a$, $2\beta$, and $N_{\rm b}$ are known. According to Table~\ref{tab:Ex2}, the values of $E[x_{k}^{2}]$ with different $2\beta$ changes little, and thus $P_{\rm modout}$ are almost the same.
\begin{table}[!h]
\caption{Mean values of $x_{k}^{2}$ ($E[x_{k}^{2}]$) during payload with different values of $2\beta$ while $a$ is fixed at 2.}
\vspace{3mm}
\centering
    \begin{tabular}{ | c | c | c | c | c |}
    \hline
   $2\beta$ & 50 & 100 & 500 & 10000\\ \hline
   $E[x_{k}^{2}]$& 1.0065 & 1.0067 & 1.0046 & 0.9996 \\ \hline
    \end{tabular}
\label{tab:Ex2}
\end{table}
\begin{table}[!h]
\caption{Power transferred through modulator with different values of $2\beta$ while $a$ is fixed at 2. $P_{\rm modoutsim}$ and $P_{\rm modout}$ corresponds to simulation result and calculation result, respectively.}
\vspace{3mm}
\centering
    \begin{tabular}{ | c | c | c | c | c |}
    \hline
   $2\beta$ & 50 & 100 & 500 & 1000\\ \hline
   $P_{\rm spay} / W$& 2.516 & 2.517 & 2.512 & 2.499 \\ \hline
   $P_{\rm modout} / W$& 2.51625 & 2.51675 & 2.5115 & 2.499 \\ \hline
    \end{tabular}
\label{tab:p}
\end{table}

\section{Conclusions}
In this paper, we proposed to encrypt power packets before sending them in a power packet dispatching system for the purpose of protecting the content of power packets. At first, we summarized the principles of the DCSK scheme referring to other researchers' work. Then partial power packet encryption method using DCSK scheme was proposed. The security of information and power of packet were discussed, seperately.  Next, we proposed to encrypt the whole power packet in order to further improve the security. The power of encrypted power packet was then examined in simulations. 
\\\indent Applying power packet encryption, the information of packet is protected from attackers. The power is protected in a certain extent from the viewpoint of the amount of stolen power. In addition, due to the principles of DCSK scheme, the power of packet after whole encryption can be rescaled. This might be a potential advantage to expand the range of power sources to meet the demand of loads.

\acknowledgments
This work was supported by Council for Science, Technology and Innovation (CSTI), Cross-ministerial Strategic Innovation Promotion Program (SIP), ``Next-generation power electronics'' (funding agency: NEDO). The authors also acknowledge to Prof. Ken Umeno and Prof. Yasuo Okabe for fruitful discussions. The author (R.T.) was partially supported by the JSPS, Grant-in-Aid for Young Scientist (B), 26820144.

\end{document}